# An Improved Version of the Symmetrical Immune Network Theory

Geoffrey W. Hoffmann[1]

An improved version of the symmetrical immune network theory is presented, in which killing is ascribed to IgM antibodies, while IgG antibodies are stimulatory. A recurring theme in the symmetrical network theory is the concept of co-selection. Co-selection is the mutual positive selection of individual members from within two diverse populations, such that selection of members within each population is dependent on interaction with (recognition of) one or more members within the other population. The theory resolves the famous I-J paradox of the 1980s, based on co-selection involving helper T cells with some affinity for MHC class II and suppressor T cells that are anti-anti-MHC class II. The theory leads to an experimentally testable prediction concerning I-J. A mathematical model that simulates IgM killing and inhibition of IgM killing by T cells is surprisingly the same as one that models a neural network.

---

[1]Department of Physics and Astronomy, University of British Columbia, Vancouver, B.C. Canada and Network Immunology Inc., 3311 Quesnel Drive, Vancouver, B.C., Canada V6S 1Z7. E-mail: hoffmann@networkimmunologyinc.com

The symmetrical immune network theory (*1-4*) is the most detailed version of immune network theory. The theory is based on Jerne's network hypothesis (*5*). In this paper the basic aspects of this theory are reviewed, and an improvement to the theory is described. The theory leads to a novel and simple HIV vaccine concept.

Many papers published in the 1970s and 1980s reported that T cells make molecules that were called specific T cell factors, and that are involved in the regulation of the production of antibodies by B cells. For example, Takemori and Tada showed rigorously that specific T cell factors are able to exert a potent specific inhibitory effect on the production of antibodies (*6*). They found that specific T cell factors have a molecular weight in the 35,000 to 60,000 dalton range, in contrast to IgG antibodies, that have a molecular weight of about 150,000. Specific T cell factors play a central role in the symmetrical network theory, and for the sake of brevity I will call them tabs. Each tab is postulated to have just one V region, in contrast to IgG molecules, that each have two V regions.

Most immunologists currently play little heed to the data supporting the existence of tabs. The main reason is that suppressor tabs were found to express a serological marker called I-J (*7*), and there is no gene for I-J at the site in the mouse major histocompatability complex where I-J had been mapped (*8*). The reasoning has been "There is no gene for I-J, therefore I-J does not exist, therefore molecules that express I-J determinants do not exist." However, as discussed below, the I-J paradox can be resolved in the context of the symmetrical immune network theory, and the data demonstrating both the existence of I-J and the existence of tabs stands.

**The symmetrical immune network theory.** The symmetrical immune network theory is based firstly on symmetrical interactions between antigen-specific cells their specific products on the one hand, and cells and molecules with complementary specificities, called antiidiotypic cells and molecules. There are three types of symmetrical interactions in the theory. Firstly symmetrical stimulation, which follows from the idea that the stimulation of lymphocytes involves cross-linking of specific receptors. For example, an antigen-specific antibody has the ability to cross-link the specific receptors of an antiidiotypic cell, and an antiidiotypic antibody has the ability to cross-link the receptors of an antigen-specific cell. Symmetrical inhibition is based on the postulate that tabs have only one V region. Then antigen-specific tabs are able block the receptors of antiidiotypic cells, and antiidiotypic tabs are able to block the receptors of antigen-specific cells. There is also a stimulatory role for tabs in the theory. Tabs bind to a receptor on non-specific accessory cells including macrophages (A cells) (*9*), and this receptor is postulated to be at least divalent. Then antigen-specific tabs on A cell surfaces can stimulate antiidiotypic lymphocytes, and antiidiotypic tabs can stimulate antigen-specific cells. Symmetrical killing involves antigen-specific IgM antibodies plus complement killing antiidiotypic cells and antiidiotypic IgM antibodies killing antigen-specific cells (*10*). In the previous model, there was linear killing by IgM and quadratic killing by IgG. In the improved model there is only linear killing by IgM, while IgG has a stimulatory role. The concept that IgM killing is more important than IgG killing is justified firstly by the fact that each IgM antibody has ten V regions as opposed to two in the case of IgG, and therefore antigen-specific IgM antibodies bind to a wider spectrum of antiidiotypic cells than is the case for antigen-specific IgG antibodies. Secondly, complement mediated killing by IgM is more efficient than complement mediated killing by IgG, since a single IgM molecule bound to a cell surface suffices to fix complement, while at least two IgG molecules bound next to each other are needed to fix complement (*11*).

Cross-linking the receptors of B cells causes the B cells to proliferate (*12*), and they need a second signal to differentiate into antibody producing cells, as reviewed in reference 2. Activation of T cells to proliferate is also dependent on a second signal cytokine (*13*). In the theory it is postulated that A cells can be activated by antigen cross-linking antigen-specific tabs bound to A cell surfaces, and the activated A cells secrete the cytokines needed for T cell proliferation and the differentiation of B cells to become plasma cells, that secrete large amounts of antibodies.

The phenomena of how the system can respond to increasing doses of antigen has been a challenge for theorists since Mitchison showed that small or large doses of an antigen can cause smaller immune responses than intermediate doses of antigen (*14*). Low dose tolerance also refers to the phenomenon that low priming doses of an antigen can result in a weak response to a subsequent otherwise immunogenic dose of antigen (*15*). This low dose tolerance phenomenon is a paradox in the context of clonal selection, without taking account of network interactions. In the context of the symmetrical network theory we can explain low and high dose tolerance as follows.

In the case of too little priming antigen, there is arming of the A cell with antigen-specific tabs, but little if any activation of the A cell by antigen. The A cells armed with antigen-specific tabs stimulate antiidiotypic cells, that secrete antiidiotypic tabs. The A cell is then armed with a mixture of antigen-specific tabs and antiidiotypic tabs, and is a catalyst for the mutual stimulation of the antigen-specific and antiidiotypic T cell populations. The system goes to a relatively unresponsive state characterized by high levels of antigen-specific and antiidiotypic T cells and tabs. The same happens for a super-optimal amount of antigen, since the activation of the A cell by antigen is inhibited by excess antigen-specific tabs, that on the A cells can again stimulate the production of antiidiotypic tabs and the proliferation of antiidiotypic T cells. Elevated and stable levels of antigen-specific and antiidiotypic T cells correspond to inhibition by antigen-specific and antiidiotypic tabs, and the suppressed state for the antigen. In summary, if the A cell is activated, the system switches to the secretion of antibodies and immunity to the antigen, while if the A cell is not activated, it switches to a non-responsive or suppressed state for the antigen.

Until now the symmetrical immune network theory has included an "immune state" in which the level of antiidiotypic cells is much less than the level of antigen-specific clones due to the killing of antiidiotypic cells by antigen-specific IgG (*1, 2*). In the 1980s Robert Forsyth and I did some experiments to determine whether we could obtain data supporting that notion (*16*). Surprisingly, we found that immunization of mice and chickens with the antigen bovine serum albumin resulted in quite the opposite. The mice and chickens both made IgG antibodies that could be affinity purified on an antigen column, and the purified antibodies had both antigen-binding and antiidiotypic specificity. In the context of the rest of the theory, these results are most simply interpreted in terms of both antigen-specific and antiidiotypic tabs on A cells stimulating B cells with the dual specificity. This is then a case of positive selection, with the fittest clones being those with antigen-specific and antiidiotypic specificities, and no evidence of a role for IgG mediated killing of clones with the complementary specificity.

**Co-selection in immune responses.** Co-selection is a term that I have introduced to denote the mutual positive selection of individual members from within two diverse populations, such that selection of members within each population is dependent on interaction with (recognition of) one or more members within the other population (*4*). Co-selection is a recurring theme in the symmetrical immune network theory. In the case of the reaction to a foreign antigen, there are two fundamentally different types of co-selection, namely co-selection without symmetry breaking and co-selection with symmetry breaking. Before being exposed to the antigen, the system is in a symmetrical virgin state with respect to the antigen. I will now describe how tolerance induction involves co-selection without symmetry breaking, while the immune response involves co-selection with symmetry breaking.

In the case of tolerance induction, symmetry is not broken because the antigen quickly becomes less important than antigen-specific and the antiidiotypic tabs on the surface of the A cells. The antigen only triggers the process. It causes co-selection of antigen-specific T cells and antiidiotypic T cells, taking the system to elevated levels of both, without any preference of one over the other. The resulting symmetry of the unresponsive state includes symmetry with respect to the diversity of antigen-specific and antiidiotypic T cells; they are equally diverse. A subsequent challenge with the antigen floods the A cell with antigen-specific and antiidiotypic tabs, increasing the number of antigen-specific and antiidiotypic T cells equally, and maintaining the symmetry of the system, while inhibiting the activation of the A cell by the antigen. This highly symmetric state remains symmetric and unresponsive to the antigen.

The induction of immunity with memory involves the production of IgG antibodies, which together with IgM antibodies rapidly clear the system of the antigen. At this stage the antigen-specific IgG antibodies are typically very diverse, but they have an average shape that is complementary to the shape of the antigen, and in aggregate they preferentially stimulate antiidiotypic cells with receptors that resemble the shape of the antigen. We again have co-selection of an antigen-specific and an antiidiotypic population, but this is co-selection with symmetry breaking, because while the antigen-specific population is diverse, the antiidiotypic cells are selected on the basis of having receptors with complementarity to as many antigen-specific cells as possible. The latter constraint results in the antiidiotypic population having very limited diversity compared with that of the antigen-specific population, and that is the asymmetric aspect. Mutual stimulation of the antigen-specific and antiidiotypic populations results in elevated levels of both populations, but the symmetry is broken with regard to amount of diversity in the two populations, and this asymmetry means that the antigen-specific population is not as tightly regulated. Challenge with the antigen then results in the relatively homogeneous antiidiotypic tabs on the A cells being a very strong stimulus for antigen-specific B cells, and causes a strong secondary IgG response.

**The Oudin and Cazenave paradox.** Oudin and Cazenave found that antibodies to multiple determinants on an antigen can express common idiotypes (*17*). The antigen was also found to induce antibodies that are not specific for the antigen at all. None of this makes sense without invoking network regulation. It can however be understood as the result of co-selection of antigen-specific and antiidiotypic clones with symmetry breaking as described above and as shown in Fig. 1. Stimulation by the antigen leads to co-selection of (a) anti-A and anti-B idiotypes and (b) antiidiotypes that are both anti-anti-A and anti-anti-B. The antiidiotypic V regions (b) are then homogeneous and emerge as the strongest antigens in the system, and with the result that they select clones (c) that are both anti-anti-anti-A and anti-anti-anti-B. All of the clones (c) express the idiotype defined by the antiidiotypic reagent (b), and some but not all of them will have anti-A or anti-B specificity, or both, or neither.

**Classic suppressor T cells.** In many cases when T cells from a mouse that has been primed with an antigen are combined with naïve cells in a recipient mouse, the mouse is suppressed for the antigen (*18*). This is the classic suppressor T cell phenomenon, in which the suppressor T cells express the CD8 marker (*19, 20*), and is not to be confused with the more recently described Treg cells (*21*). The data supporting the existence of these suppressors is readily understood in terms of co-selection of antigen-specific cells and antiidiotypic T cells, that are present in the naïve and suppressed populations. The co-selection occurs via tabs adsorbed to A cells. This type of suppression is important, but these suppressor T cells typically express I-J, and again due to the I-J paradox their very existence is currently widely ignored. They are barely mentioned in current immunology textbooks (*22*), in spite of having been described in many papers. We therefore now need to turn to I-J, and describe how the paradox can be resolved in the context of the symmetrical immune network theory.

**The I-J phenomenon.** The I-J phenomenon was discovered in mice, and maps to the MHC class II region of the major histocompatability complex. The MHC class II genes are $A_\alpha$, $A_\beta$, $E_\beta$ and $E_\alpha$, in that order. In 1976 two groups independently discovered a new phenomenon they called I-J (*7, 23*). Mice were immunized with lymphoid cells from strains that differed or were believed to differ in part or all of their MHC. The mice made antibodies that bound specifically to suppressor tabs (*7*) and to suppressor T cells (*23*). These were called anti-I-J antibodies. So the tabs and suppressor T cells that play a role in suppressing immune responses express I-J shapes, as defined by the anti-I-J antibodies. Careful experiments using many inbred strains of mice seemed to show that I-J is encoded within the MHC of the mouse between the $E_\beta$ and $E_\alpha$ genes.

From the perspective of the symmetrical immune network theory, this was exciting, because the phenomenon of suppression was clearly important, suppression seemed to be mediated by tabs, and tabs play a central role in the symmetrical network theory. Furthermore, I-J was not a fly-by-night finding. About 1000 papers were published with I-J in the title.

However, in 1982 a problem emerged. DNA sequencing revealed that there was no gene that could encode I-J gene in the MHC class II region where I-J had been mapped (*8, 24*). This was the I-J paradox. Much effort was devoted to resolving this problem, but the solution was slow in coming.

With time, many immunologists threw out the baby with the bathwater. In spite of thousands of papers on antigen-specific suppression mediated by T cells in mixing experiments, and the many papers on suppressor tabs that express I-J, these findings were relegated by most immunologists to history.

This reaction was and is however inappropriate. When we have a paradox, it is telling us that we are not understanding something. It does not mean we should throw out a huge amount of information and call it all nonsense. As long as the original papers that defined I-J stand, we need to persist with trying to find a solution.

A partial solution to the I-J paradox came with the findings that the I-J phenotype of suppressor T cells depends on the MHC environment of the T cells during ontogeny (*25*, *26*). In the context of network theory, it is reasonable to consider the possibility that I-J determinants are V region determinants on suppressor T cells, that are selected such that the V regions have complementarity to helper T cell idiotypes, that in turn are selected to have some complementarity to MHC class II. In other words, I-J determinants are anti-anti-MHC class II determinants. Then there is the question of whether the topology of connections between anti-MHC class II helper T cells and anti-anti-MHC class II suppressor T cells is divergent as shown in Fig. 2a, or convergent as shown in Fig. 2b. In the context of the symmetrical network theory, suppressor T cells can be expected to have high network connectance, which is more consistent with the topology of Fig. 2b than that of Fig. 2a. Since the interaction between the helper T cells and the suppressor T cells is symmetrical, the suppressor T cell population is co-selected with the helper T cell population, with each suppressor T cell being selected on the basis of having V regions with complementarity to as many helper T cells as possible. At the same time, the helper T cells are selected to have complementarity not only to MHC class II, but also to the suppressor T cell V regions. The helper T cell repertoire is then stabilized by the suppressor T cells and vice versa. A corny analogy is that the suppressor T cell V regions are like the centre-pole of a tent, and are homogeneous, while the helper T cell V regions are like a canvas that stabilizes the centre-pole, and are more diverse. For a given set of MHC class II antigens, there can be more than one combination of mutually stabilizing helper T and suppressor T cell populations.

It follows that mice which are genetically identical can express different I-J phenotypes. The initial conditions during embryogenesis can be expected to play a role in determining the nature of the I-J shapes that emerge, helping to ensure that the offspring of mice with a given I-J phenotype have the same I-J phenotype.

While MHC class II has a particularly strong impact on the repertoire of helper T cells, non-polymorphic self antigens can reasonably be expected to also play a role in the selection of the helper T cell repertoire and that of the co-selected suppressor T cells. We cannot expect to see the impact of such non-polymorphic self antigens in the types of experiments that show the impact of the polymorphic MHC self antigens.

This resolution of the I-J paradox leads to a simple experimentally testable prediction as follows. In an extension of work by Binz and Wigzell (*27*), Cooper-Willis, Chow and I found that the immune response of a mouse of strain A to immunization with lymphoid cells of a strain B is complementary to the immune response of the strain B to immunization with lymphoid cells of the strain A. We called this phenomenon second symmetry (*28*). Consider the case that the strains A and B differ only in their MHC genes. Then the B strain antigens to which the A mice respond include

1. Conventional $MHC^B$ antigens
2. B strain anti-A receptors
3. $I-J^B$ receptors = anti-anti-B

and the A strain antigens to which the B strain mice respond include

1. Conventional $MHC^A$ antigens
2. A strain anti-B receptors
3. $I-J^A$ receptors = anti-anti-A

The immune response of strain A to strain B cells ("A anti-B") then includes

1. anti-$MHC^B$
2. anti-anti-A
3. anti-$I-J^B$ = anti-anti-anti-B

and the immune response of strain B to strain A cells ("B anti-A") includes

1. anti-$MHC^A$
2. anti-anti-B
3. anti-$I-J^A$ = anti-anti-anti-A

Absorption of the A anti-B serum with strain B lymphoid cells removes the anti-$MHC^B$ and anti-$I-J^B$, leaving anti-anti-A. Then anti-$I-J^A$ antibodies are predicted to bind specifically to the remnant anti-anti-A present in A anti-B serum absorbed with B strain lymphoid cells, and anti-$I-J^B$ antibodies are likewise predicted to bind specifically to the anti-anti-B present in B anti-A serum that has been absorbed with A strain cells.

**A mathematical model that includes IgM and T cells.** I now describe a mathematical model that simulates the autonomous kinetics of the central components of the system, without including the antigen. Firstly, an appropriate model for a large number $N$ of interacting IgM producing B cells has the form

$$\frac{dx_i}{dt} = 1 - x_i \sum_{j=1}^{N} \beta_{ij} x_j \qquad (1)$$

The number of IgM secreting cells with specificity $i$ is $x_i$. The first term on the right simulates a constant influx of cells of specificity $i$ from the bone marrow. For simplicity this is taken to be the same for all $N$ clones, and can be given the value 1 for all clones, based on appropriate scaling of the variables $x_i$. The other term simulates killing of the B cells of clone $i$ by IgM molecules that have specificities that are complementary to that of clone $i$. The matrix $\beta_{ij}$ models the affinity of clone $i$ V regions to clone $j$ V regions. This matrix is symmetric, and has a connectance $C$ defined as the fraction of non-zero terms. We consider the case that the matrix is otherwise random. Equation (1) is a qualitatively unusual differential equation, in that for the system to have a stable steady state it needs a minimal level of complexity (*3*). High $C$ and high $N$ correspond to a high level of complexity, while low $C$ and low $N$ correspond to a low level of complexity. When this system is stable, it has a single attractor, meaning it converges to the same stable steady state regardless of initial conditions. The threshold level of connectance for the system to be stable is low, and corresponds to the system being stable providing each clone interacts with approximately two or more other clones.

      We now add natural death to the equation, together with the inhibition by tabs of IgM killing, in the case of specificities for which there is a high level of T cells. Natural death is modeled by adding a term $-x_i$ in the differential equation. In the case of specificities for which there is a high level of T cells, the inhibition of complement plus IgM mediated killing depends on both the level of the clone $i$ and the level of clones that have complementarity to clone $i$. In this mathematical model the subscript $i$ refers to a shape $i$. There are IgM secreting cells that have V regions with shape $i$, and there are also T cells with V regions with shape $i$, in addition to T cells and IgM molecules with V regions that have complementarity to the shape $i$. The catalyst role of the A cell in the mutual stimulation of T cells means that on a short time scale the system evolves towards the level of T cells for each specificity being kept at approximately the level of T cells with complementary specificities. The role of tabs as inhibitors of IgM killing is then modeled with a term that depends on $x_i^2$, since $x_i^2$ models the amount of mutual stimulation of the two sets of T cells via the A cell.

The model for the IgM network including natural death and inhibition of killing by T cells is then

$$\frac{dx_i}{dt} = 1 - x_i - \frac{x_i \sum_{j=1}^{N} \beta_{ij} x_j}{1 + \alpha x_i^2} \qquad (2)$$

where $\alpha$ is a constant. The stable states of this system are obtained by setting the left hand side equal to zero, giving a cubic equation in $x_i$, with $\sum_{j=1}^{N} \beta_{ij} x_j$ as a parameter. For a given value of the parameter, the cubic equation can have three solutions, of which two are stable steady states for $x_i$, and the system as a whole can have up to approximately $2^N$ stable steady states. The differential equation (2) simulates the autonomous dynamics of the system, without attempting to include the antigen. The switching caused by the antigen activating the A cell to produce second signal lymphokines is likewise not explicitly included in this mathematical model, and can be understood within this framework as explained in words above.

**Regulation by IgG and IgM in the mathematical model.** When there is an immune response with the production of IgG, the IgG stimulates T cells with V regions that are complementary to the IgG. These T cells make tabs that bind to A cells, that then in turn stimulate T cells with V regions that are similar to those of the IgG. In this way, a combination of antigen and antigen-specific IgG down-regulates the immune response to the antigen, and terminates it. In the mathematical model, this takes place by IgG and the antigen together effectively causing an increase in the $\alpha x_i^2$ term and a decrease in IgM mediated killing, limiting the extent to which the symmetry between antigen-specific and antiidiotypic clones is broken.

Henry and Jerne showed that administration of IgG antibody specific for sheep red blood cells (SRBC) one to two hours prior to immunization of mice with SRBC results in a profound suppression of the immune response to the SRBC (*29*). The IgG antibodies together with the antigen constitute a dual stimulus, causing co-selection of antigen-specific and antiidiotypic T cells, leading to the suppressed state with elevated levels of both. The idiotypic-antiidiotypic symmetry is broken less than in the case of only the antigen being administered, and the system goes to a relatively suppressed state. Henry and Jerne also showed that IgM specific for SRBC given just prior to SRBC causes an enhanced immune response. This can be ascribed to the IgM killing antiidiotypic T cells, and thus assisting in the breaking of symmetry between antigen-specific and antiidiotypic cells.

**On the analogy with the brain.** There has been a powerful interplay between the development of immune system theory and neural network theory. In his original network hypothesis paper Jerne emphasized the similarity between the immune system and the brain (*5*). These similarities include the large number and diversity of cells, permitting appropriate responses to be made to an enormous variety of stimuli. In both cases the cells are functionally connected to each other as a network. Both systems exhibit memory, which can last for years. This aspect distinguishes the immune system and the brain from all other physiological systems. In both systems the acquired memories are not passed on to subsequent generations, even though it would be advantageous for this to be the case. Furthermore, both systems have a profound sense of self. I utilized the similarity between the immune system and the brain to formulate a neural network model in which neurons exhibit hysteresis (*30*). This system has close to $2^N$ stable steady states, and is capable of learning through interacting with its environment without changes in synaptic connection strengths (*31*). The equation for the neural network model is the same as the symmetrical immune network equation (2) above. The two systems consist of very different components, but they are modelled by the same differential equation. This suggests that the number of ways of constructing adaptive biological systems, that exhibit both a large number of stable steady states and the ability to learn, is severely limited, and nature has surprisingly found two different ways to construct systems using very different building blocks, and yet are based on the same differential equation.

**Conclusion.** The constraint of finding explanations for phenomena that do not make sense in the context of basic clonal selection, meaning clonal selection without taking account of idiotypic network interactions, is a powerful constraint for the formulation of a theory of regulation of the adaptive immune system. In this paper we have shown that an improved version of the symmetrical immune network theory resolves important paradoxes, including low dose tolerance, the Oudin-Cazenave paradox and the I-J paradox among others. It also accounts for a new phenomenon that we have discovered, namely MHC restriction of V-V interactions in serum IgG. A lot of data on suppressor T cells and tabs is currently being neglected, buried in forgotten literature. Immunologists will benefit from revisiting these phenomena and addressing the question of whether the symmetrical network theory provides the best way for understanding the adaptive immune system. This is particularly important in light of the fact that the theory has potential for the development of a preventive HIV vaccine.

**Fig. 1.** A co-selection model that accounts for the Oudin-Cazenave paradox. The antigen-specific set αA (anti-A) and αB (anti-B) includes clones that are specific for various antigenic determinants of the antigen, here A and B, while the ααA and ααB anti-idiotypic set is selected to have specificity for as many antigen-specific clones as possible. The emergent antiidiotypic set is the strongest antigen in the system, and induces clones that by definition are αααA and αααB. These include clones that are anti-A or anti-B, clones that are both anti-A and anti-B, and clones that are neither anti-A nor anti-B.

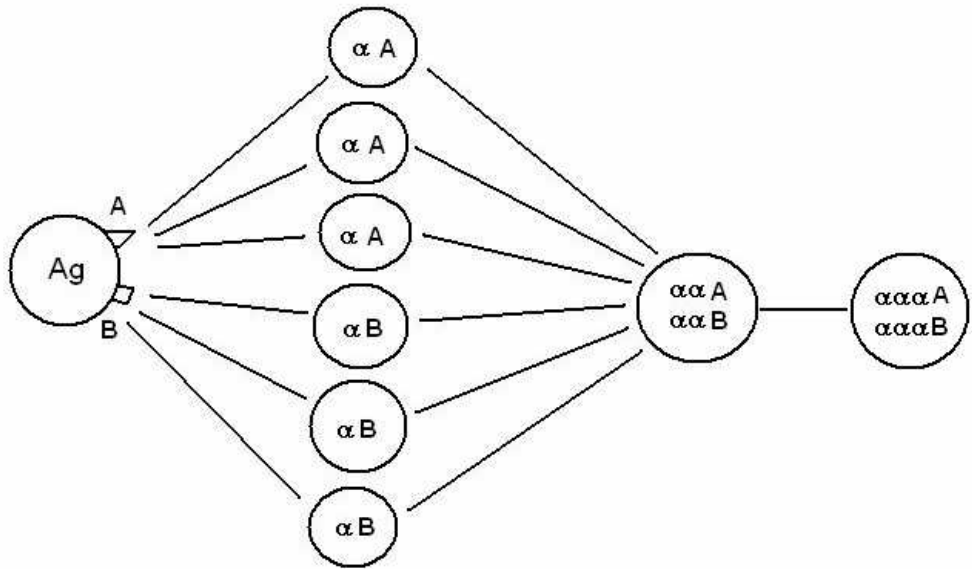

**Fig. 2.** Network topologies for clones that relate directly or indirectly to MHC class II self antigens. **a**. A divergent topology. **b**. A convergent topology that is the basis for accounting for the I-J paradox.

**a**

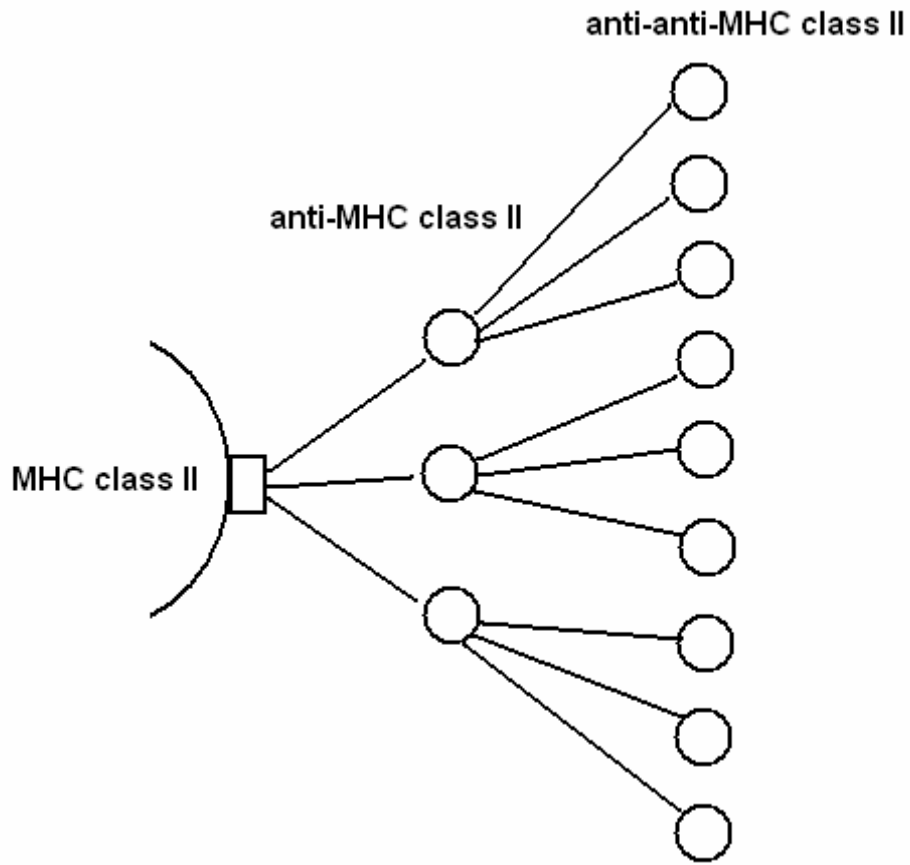

b

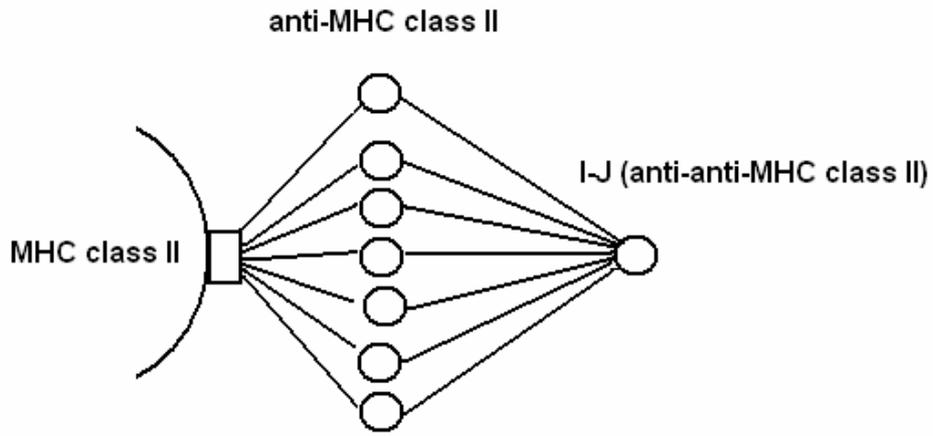